%% file: main.tex
\pdfoutput=1

\documentclass[11pt]{article}

\usepackage[]{acl}

\usepackage{times}
\usepackage{latexsym}

\usepackage[T1]{fontenc}

\usepackage[utf8]{inputenc}

\usepackage{microtype}

\usepackage{inconsolata}
\usepackage{tikz}
\usepackage{amsmath}
\usepackage{enumitem}
\usepackage[english]{babel}
\usepackage{bbding}
\usepackage{multirow}
\usepackage{float}
\usepackage{amssymb}
\usepackage{graphicx}
\usepackage{newfloat}
\usepackage{listings}
\usepackage{caption}
\usepackage{stfloats}
\usepackage{booktabs}
\usepackage{subfigure}
\usepackage{xcolor}
\usepackage{multirow}
\usepackage{listings}
\usepackage{algorithm}
\usepackage{algorithmicx}
\usepackage{algpseudocode}
\usepackage{hyperref} 
\usepackage{xspace}
\usepackage{cleveref}
\usepackage{bbm}

\newtheorem{definition}{Definition}

\newcommand{\ours}{\textsc{Buzzer}\xspace}
\newcommand{\task}{CMI\xspace}

\title{Code Membership Inference for Detecting Unauthorized Data Use in\\Code Pre-trained Language Models}

\author{Sheng Zhang,\,\, Hui Li\thanks{$\quad$Corresponding Author.},\,\, Rongrong Ji \\
        Key Laboratory of Multimedia Trusted Perception and Efficient Computing\\ Ministry of Education of China, Xiamen University \\
        \texttt{sheng@stu.xmu.edu.com},\,\, \texttt{\{hui,rrji\}@xmu.edu.com}
        }

\begin{document}

\maketitle

\begin{abstract}
Code pre-trained language models (CPLMs) have received great attention since they can benefit various tasks that facilitate software development and maintenance. However, CPLMs are trained on massive open-source code, raising concerns about potential data infringement. This paper launches the study of detecting unauthorized code use in CPLMs, i.e., Code Membership Inference (CMI) task. We design a framework \textsc{Buzzer} for different settings of CMI. \textsc{Buzzer} deploys several inference techniques, including signal extraction from pre-training tasks, hard-to-learn sample calibration and weighted inference, to identify code membership status accurately. Extensive experiments show that CMI can be achieved with high accuracy using \textsc{Buzzer}. Hence, \textsc{Buzzer} can serve as a \task tool and help protect intellectual property rights. The implementation of \textsc{Buzzer} is available at: \url{https://github.com/KDEGroup/Buzzer}.
\end{abstract}

\input{tex/introduction.tex}
\input{tex/background.tex}

\input{tex/methodology.tex}
\input{tex/experiment.tex}

\input{tex/conclusion.tex}
\input{tex/limitations.tex}

\section*{Acknowledgments}
This work was partially supported by National Science and Technology Major Project (No. 2022ZD0118201), Natural Science Foundation of Xiamen, China (No. 3502Z202471028) and National Natural Science Foundation of China (No. 62002303, 42171456).

\bibliography{ref}

\end{document}

%% file: tex/introduction.tex

\section{Introduction}

Recently, various code pre-trained language models (CPLMs) like CodeBERT~\cite{CodeBERT} and Code Llama~\cite{abs-2308-12950} have sprung up and shown strong capabilities.
CPLMs are pre-trained over massive code data that is publicly available in platforms like GitHub and StackOverflow. 
Then, CPLMs can be fine-tuned or directly used for code-related tasks like code refactoring~\cite{LiuWWXWLJ23} and code search~\cite{WangJLYX0L22} even when the downstream tasks do not have much data, reducing the intellectual burden of developers and facilitating software development and maintenance

However, using code data to train CPLMs may cause patent infringement and legal violations. 
GitHub recently introduced a programming tool Copilot\footnote{\url{https://github.com/features/copilot}}.
Copilot is powered by OpenAI Codex, a GPT based CPLM.
However, Copilot has faced allegations of violating open-source licenses~\cite{InfoQ22} since it is trained on code that may be collected from open-source projects. 
Although it is still under debate whether using open-source code to train CPLMs causes intellectual property infringement, the lawsuit has alerted researchers and companies who work on CPLMs: code data from open-source projects is \emph{not free} training data.

To help protect the intellectual property rights on code data, this paper studies a new task named \underline{C}ode \underline{M}embership \underline{I}nference (\task) for CPLMs which identifies whether a well-trained CPLM used a certain code snippet as its training data. 
A \task method for CPLMs can serve as a tool to detect unauthorized data use and provide potential evidence when a lawsuit similar to Copilot's case is filed in the future.
We propose a \task framework \ours for detecting unauthorized data use in different settings of \task.
The contributions of this work are summarized as follows:
\begin{enumerate}
    \item We define two levels of inference for CPLMs: white-box inference and black-box inference. They have different knowledge w.r.t. CPLMs and training data. White-box inference is hard to achieve, but it can help us understand the upper bound of the accuracy of \task, while black-box inference is more likely to succeed in practice.

    \item For the two settings, our proposed \ours applies various inference techniques, including signal extraction from pre-training tasks, hard-to-learn sample calibration and weighted inference, to identify code membership status accurately. 

    \item We have conducted \task on representative CPLMs. Experimental results show that \ours can achieve promising accuracy. Additionally, we find that the accuracy of \ours in black-box inference is not much worse than that in white-box inference, showing that \task can be achieved with high accuracy in practice.

\end{enumerate}

%% file: tex/background.tex

\section{Related Work}
\label{sec:Background}

\subsection{Code Pre-trained Language Model}
\label{sec:CPLM}

Prevalent CPLMs~\cite{CodeBERT,GraphCodeBERT,UniXcoder} typically adopt a multi-layer Transformer architecture~\cite{selfAttention} with $N$ Transformer blocks (we call them hidden layers in this paper).
Given a code snippet $c$, CPLM encodes it into high-dimension representation vectors.
Before feeding $c$ into the CPLM, 
it is natural to tokenize $c$ into a series of tokens $\{c_1\cdots c_n\}$.
Then, tokens will be encoded by the CPLM into representation vectors $\{\mathbf{h}_1,\cdots,\mathbf{h}_n\}$ that can be further used in downstream code-related tasks.
Note that, CPLMs can also encode the corresponding descriptions $\{o_1\cdots o_m\}$ of $c$ (e.g., method comment) written in natural language into token representations $\{\mathbf{r}_1,\cdots,\mathbf{r}_m\}$~\cite{CodeBERT}.

According to their model architectures, recent works can be categorized into three types:
\begin{itemize}
    \item \textbf{Encoder Based Models.}
    Encoder-only CPLMs typically follow the design of BERT~\cite{abs-1907-11692}.
    CodeBERT~\cite{CodeBERT} uses masked language modeling and replaced token detection tasks for pre-training. 
    GraphCodeBERT~\cite{GraphCodeBERT} models code data from a structural perspective and it uses edge prediction and node alignment as the pre-training tasks. 

    \item \textbf{Decoder Based Models.}
    Decoder based CPLMs only utilize multi-layer transformer decoders, 
    and they are known for their enhanced generalization capabilities in generative tasks~\cite{abs-2305-15525}. 
    IntelliCode~\cite{IntelliCode} and CodeGPT~\cite{CodeGPT} follow the objective of GPT-2, 
    employing the next token prediction task for pre-training.
    Based on Llama2~\cite{abs-2307-09288}, CodeLlama~\cite{abs-2308-12950} expands the model input length to 16k tokens and performs the pre-training task of fill-in-the-middle~\cite{abs-2207-14255}. 
    DeepSeek-Coder~\cite{guo2024deepseek} is pre-trained on a vast dataset containing 87 programming languages with dependency parsing and repo-level deduplication. It undergoes training for both the next token prediction and fill-in-the-middle tasks.

    \item \textbf{Encoder-Decoder Based Models.}
    Encoder-Decoder based CPLMs contain both encoder and decoder in transformer, 
    and they perform well for both understanding and generation tasks. 
    Jiang et al. propose TreeBERT~\cite{TreeBERT}, 
    which utilizes tree structure of abstract syntax trees (ASTs) and models them as a set of composition paths to enhance the understanding of code data. 
    SPT-Code~\cite{SPTCode} leverages ASTs to enhance semantic representation. 
    It improves the generation ability of CPLMs by setting up special pre-training tasks, including Code-AST prediction, Masked Sequence to Sequence (MASS)~\cite{MASS}, and method name generation. 
    CodeT5~\cite{0034WJH21} employs a unified framework to seamlessly support both code understanding and generation tasks, and it allows multi-task learning. 
    UniXcoder~\cite{UniXcoder} utilizes prefix adapters to 
    control the model behaviors and leverages multimodal data for 
    enhancing code comprehension and code generation tasks. 
\end{itemize}

\subsection{Membership Inference}

Membership Inference (MI)~\cite{HuSSDYZ22} aims to ascertain whether a given data record is part of a particular dataset used to train a specific model. 

Shokri et al.~\cite{ShokriSSS17} study membership inference by utilizing multiple shadow models to mimic the target model.
Following Shokri, Salem et al.~\cite{Salem0HBF019} relax the restrictions by reduce the number of shadow models and perform membership inference with less knowledge of member data
Yeom et al.~\cite{YeomGFJ18} investigate the role of overfitting in membership inference for popular machine learning algorithms. 
Li et al.~\cite{LabelOnlyMembership} perform membership inference by only accessing the 
final predicted label, instead of acquiring the logits or probabilities.

For language models, Song et al.~\cite{SongR20} study membership inference for word embedding models by calculating the average similarity in a sliding window. Mahloujifar et al.~\cite{mahloujifar2021membership} leverage the semantic relationships preserved by word embeddings to 
identify special word pairs. Jagannatha et al.~\cite{jagannatha2021membership} investigate the risk of training data leakage in clinical models. Mireshghallah et al.~\cite{MireshghallahGU22} introduce a reference model and give the determination based on the likelihood ratio threshold.
\task is a specific type of membership inference (MI)~\cite{HuSSDYZ22}.
Yang et al.~\cite{abs-2310-01166} studies the code membership inference task for auto-regressive models.
Their work is closely related to ours.
Differently, our method can be applied to CPLMs with other architectures in addition to auto-regressive CPLMs.

%% file: tex/methodology.tex

\section{Code Membership Inference in CPLMs}
\label{sec:method}

\subsection{Task Definition}
\label{sec:td}

We first give the definition of the \task task:
\begin{definition}[Code Membership Inference]
\label{def:cmi}

    Given a tokenized code snippet $\{c_1, c_2, \cdots, c_n\}$, a tokenized corresponding natural language descriptions $\{r_1, r_2, \cdots, r_m\}$ and the target CPLM $\mathcal{M}$, the adversary adopts an inference model to determine whether $c$ is in the training data of $\mathcal{M}$. 
\end{definition}

Instead of giving ``hard prediction'', the inference model can output a continuous \emph{confidence score} indicating the probability of $c$ being the code member data. Then, the adversary uses a threshold $\theta$ to yield the prediction:
\begin{equation}
\label{eq:threshold}
\mathcal{A}(c)=\mathbbm{1}\left[\mathcal{I}(c)>\theta\right],
\end{equation}
where $\mathbbm{1}$ is the indicator function, $\theta$ is a chosen threshold, $\mathcal{I}(\cdot)$ is the inference model that produces the confidence score, and $\mathcal{A}(\cdot)$ is the membership indicator.

\begin{figure*}[t]
    \centering
    \includegraphics[width=1\linewidth]{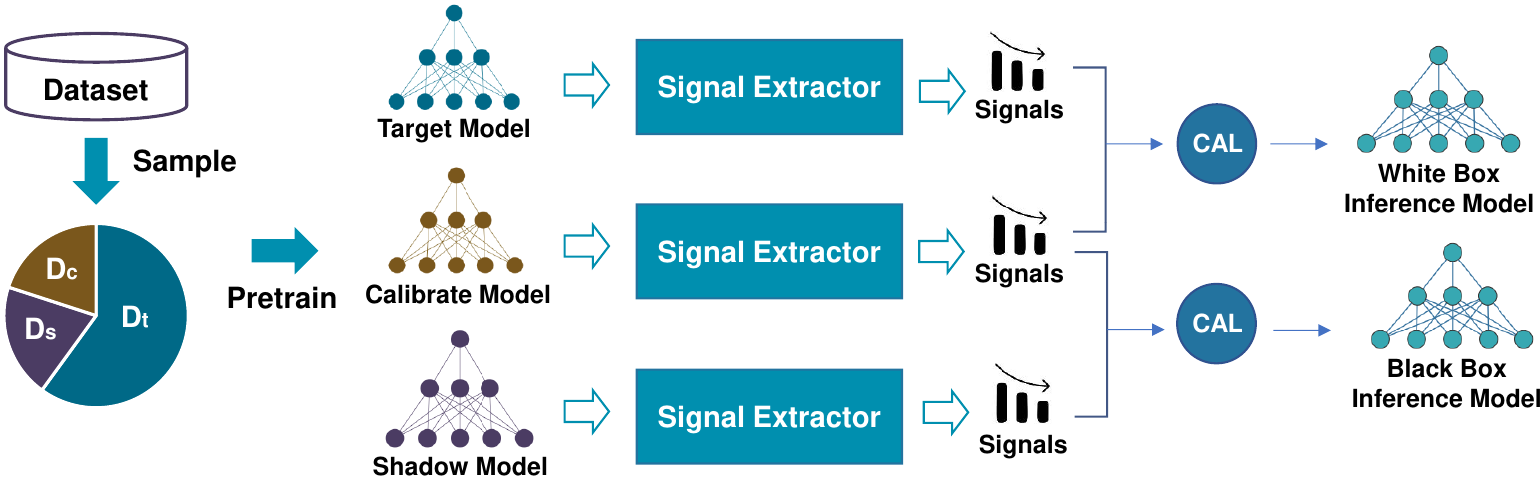}
    \caption{Overview of our \ours framework. 
    Firstly, it samples three disjoint datasets, $D_t$, $D_s$ and $D_c$, to construct target, shadow and calibrated models, respectively. 
    After that, it extracts model signals with calibration and trains white-box and black-box classifiers for \task.
    }
    \label{fig:framework}
\end{figure*}

\subsection{Knowledge Level}

The knowledge of the adversary on $\mathcal{M}$ is critical to the success of \task.
We define two inference settings with different knowledge levels:
\begin{enumerate}
    \item \textbf{White-Box Inference:} 
    The adversary has complete knowledge of $\mathcal{M}$ (e.g., model architecture, training objectives, and the trained model parameters). 
    Moreover, the adversary can access a considerable amount of the training data, converting the problem into a supervised classification problem. \
    {(e.g., 70\%)}
    In practice, this is hard to achieve from the outside model provider.

    \item \textbf{Black-Box Inference:}
    The adversary knows the core architecture (e.g., Transformer) and pre-training objectives of $\mathcal{M}$.
    Such information is typically available via public technical reports (e.g., technical reports of CodeLlama and CodeT5 are publicly available).
    Hence, compared to white-box inference, black-box inference is a more practical setting of \task.

\end{enumerate}

\subsection{Our Proposed \ours}
\label{sec:techniques}

This section illustrates the details of \ours.
As depicted in Fig.~\ref{fig:framework}, \ours is designed for handling both white-box and black-box settings of \task.
First, it samples disjoint datasets to construct target, shadow and calibration models. 
Then, it extracts model signals with calibration and trains white-box and black-box classifiers.

\subsubsection{Overview of Two Types of \task}
\label{sec:twoinference}

\vspace{5pt}
\noindent{\textbf{White-Box Inference.}} Taking advantage of the prior knowledge on the considerable amount of training data, the adversary can train an inference model to infer membership status.
The adversary can mix known code member data and other code data (code non-member data) that is very unlikely to be the code member data to construct the inference model's training data. 
Code non-member data can be sampled from the population of the dataset $\mathcal{D}$ that are not included in the description of the training data sources of $\mathcal{M}$. This way, white-box inference becomes a binary classification problem.

The next question is how to define the behavior of $\mathcal{M}$ w.r.t. a certain code snippet $c$.Recall that the input tokens of $c$ are first encoded by the embedding layer and the embeddings are passed to the first hidden layer. 
There are multiple hidden layers in a CPLM, 
and each of them applies a non-linear function on the inputs from preceding layer.
The overall forward propagation process can be described as follows: 
\begin{equation}
    \label{eq:clm}
    \mathbf{H}_{i + 1} = \text{Layer}_{i}\left( \mathbf{H}_{i} \right),
\end{equation}
where $\mathbf{H}_{i}$ refers to the output of the $i$-th hidden layer and $\text{Layer}_{i}(\cdot)$ indicates the $i$-th hidden layer.
Hidden layers are key components that enable the CPLM to understand code data. 
Therefore, after feeding $c$ to $\mathcal{M}$, we can regards the outputs of hidden layers as the behavior of $\mathcal{M}$.

In white-box inference, 
\ours first sample a target dataset $D_t$ and a calibration dataset $D_c$ to construct a target model and a calibration model (see Sec.~\ref{sec:cal}), respectively.
For an interested code snippet $c$, a signal extractor (see Sec.~\ref{sec:signal}) undertakes several pre-training tasks of the target CPLM's on the target model and the calibration model to extract signals.
These signals will then be fed into a inference model to derive the \task outcome.

\vspace{10pt}
\noindent{\textbf{Black-Box Inference.}} In black-box inference, the adversary lacks access to the $\mathcal{M}$'s member data, raising a challenge for \task: the adversary lacks labeled member and non-member data for supervised binary classification as in white-box inference. 

To overcome this problem, \ours samples a shadow dataset $D_s$ to train a shadow model.
The shadow model is designed to imitate $\mathcal{M}$ with similar structure and training algorithms. 
The adversary knows the member data (i.e., $D_s$) and non-member data of the shadow model.
Therefore, the adversary can use the shadow model to replace the target model in the black-box setting and infer code member status.
For an interested code snippet $c$, the signal extractor undertakes several pre-training tasks of the target CPLM's on the shadow model and the calibration model to extract signals.
These signals will then be fed into the inference model for the \task task.

\vspace{10pt}
\noindent\textbf{Difference between White-box Inference and Black-box Inferencec:} The primary distinction lies in the white-box's ability to access the training data of $\mathcal{M}$, while the black-box setting lacks such access. During the training phase, the white-box inference employs both member and non-member data of $\mathcal{M}$ to train its inference model, whereas the black-box inference utilizes member and non-member data of the shadow model. During the testing phase, both white-box and black-box settings conduct tests on member and non-member data of $\mathcal{M}$. More details can be found in~\ref{Experiment:data}.

\subsubsection{Signal Extractor}
\label{sec:signal}
The signal extractor captures CPLM's behavior when encountering member or non-member data.
It extracts signals, which serves as input features to subsequent inference models, from a code snippet.

\begin{figure}[t]
    \centering
    \includegraphics[width=1\linewidth]{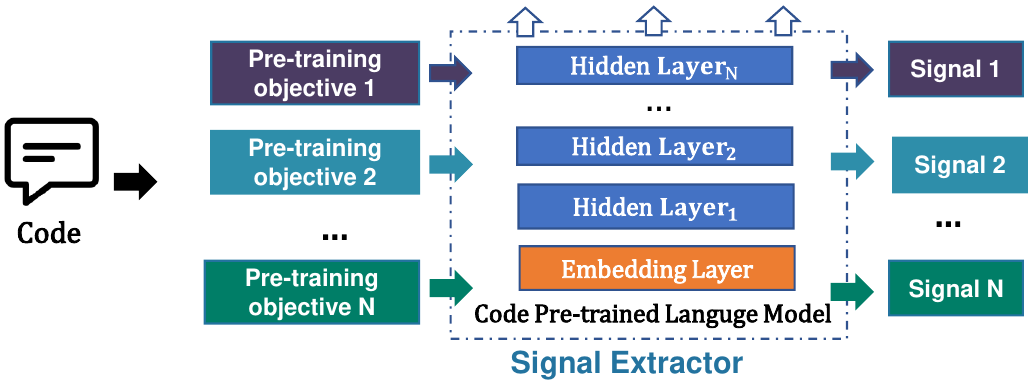}
    \caption{Overview of the signal extractor. It returns signals w.r.t. the pre-training tasks.}
    \label{fig:SignalExtractor}
\end{figure}

One question is how to define the signal of CPLMs, which should be highly correlated with both member samples and non-member samples. 
An intuitive approach is using task-specific loss values as signals. 
As the CPLM has captured member data well, encountering member data could results in lower loss of CPLM. 
Likewise, the loss values for non-member data generally turn to be lower. 

Fig.~\ref{fig:SignalExtractor} depicts the design of the signal extractor. 
For a code snippet, the signal extractor undertakes several pre-training tasks, which are consistent with the target CPLM's original pre-training tasks, on the target/shadow/calibration model. 
Through these tasks, we acquire the loss values w.r.t. input code snippet which serves as the signals. 
These signals will be used as input features for the subsequent inference model.

\subsubsection{Calibration Model}
\label{sec:cal}

Non-member samples can still produce strong signals indicating a high probability of being member data, whereas a member sample may yield the opposite result, since they are over-represented or under-represented in the data distribution~\cite{WatsonGCS22}. 
Hence, we design a calibration model to tackle such hard-to-learn samples. 
The calibration model adopts the same model architecture as the target CPLM.
The calibration model is utilized to determine sample difficulty and it is pre-trained using a dataset, which is disjoint from both the shadow model dataset and the target model dataset, and the same pre-training tasks as the target CPLM.

The hard-to-learn calibration process (``CAL'' in Fig.~\ref{fig:framework}) is illustrated in Eq.~\ref{eq:cal}, where ${signal}^{cal}_{i}$ represents the final calibration value of $i_{th}$ sample and ${signal}^{t}_{i}$ denotes the signal of the target model (white-box inference) or the shadow model (black-box inference), ${signal}^{c}_{i}$ indicates the signal of the calibration model, and $\epsilon$ prevents division by zero.
\begin{equation}
    \label{eq:cal}
    signal^{cal}_{i} = \frac{signal^{t}_{i}}{signal^{c}_{i}+\epsilon}.
\end{equation}

\subsubsection{Weighted Inference Model}

The inference model incorporates multiple signals from the signal extractor to generate the final prediction score. 
A higher score assigned by the inference model to a code snippet indicates higher likelihood of belonging to the target model's member data. 
The inference model is designed to take signals from two types of pre-training tasks:
\begin{itemize}
\item \textbf{Signals from generative pre-training tasks}: The signals generated from generative pre-training tasks (e.g., Bimodal Dual Generation, BDG)~\cite{0034WJH21}. For these signals, we utilize a one-layer self-attention based network with 12 heads to learn the feature.

\item \textbf{Signals from non-generative pre-training tasks}: The signals generated from non-generative pre-training tasks (e.g., Masked Language Modeling, MLM)~\cite{CodeBERT}. For these signals, we utilize a three-layer multi-layer perceptron based network to learn the feature.
\end{itemize}

Consider a target CPLM with pre-training tasks such as MLM and BDG. 
Initially, signals are extracted from the loss values of the pre-training tasks. 
Subsequently, the inference model adopts two different sub-networks, which take different types of signals as input, to predict the confidence scores. 
For signals generated from generative pre-training tasks (e.g., BDG), sub-networks with the self-attention mechanism are employed. 
For signals from non-generative pre-training tasks (e.g., MLM), sub-networks with three-layer MLP with ReLU are utilized. 
Finally, \ours weights the confidence scores of different sub-networks as the final confidence score in Eq.~\ref{eq:threshold}.

The loss function of the inference model is illustrated in Eq.~\ref{eq:lossFunction}, which is designed to maximize the model output of a member sample and minimize that of a non-member sample.
\begin{equation}
    \label{eq:lossFunction}
    \mathcal{L}(\Theta, m, n) = \alpha - C(m) + C(n),
\end{equation}
where $m$ and $n$ represent the member and non-member sample, respectively. $C(m)$ is the output of inference model and $\alpha$ is a hyper-parameter.

%% file: tex/experiment.tex

\section{Experiments}
\label{sec:Experiment}

\subsection{Settings}

\subsubsection{Evaluation Metrics}
We adopt Area Under the Curve (AUC) as the main evaluation metrics, which assesses a model's capability to differentiate between positive and negative samples.
AUC is widely used in evaluating MI~\cite{LabelOnlyMembership,MireshghallahGU22,ZhangRWRCHZ21,WangH0RCLRR22}.
We also consider True Positive Rates (TPR) at low False Positive Rates (FPR) as the evaluate metric~\cite{CarliniCN0TT22}. 
Specifically, we compare TPR values of different methods when the target FPR values are low (1\%, 0.1\% and 0.01\%).

\subsubsection{CPLMs}

We choose four representative CPLMs as target models, including CodeBERT\footnote{\url{https://github.com/microsoft/CodeBERT}}~\cite{CodeBERT}, CodeT5\footnote{\url{https://github.com/salesforce/CodeT5}}~\cite{0034WJH21}, DeepseekCoder\footnote{\url{https://huggingface.co/deepseek-ai/deepseek-coder-1.3b-base}}~\cite{guo2024deepseek} and CodeLlama\footnote{\url{https://huggingface.co/codellama/CodeLlama-7b-hf}}~\cite{abs-2308-12950}.
CodeBERT is a BERT based bimodal CPLM.
CodeT5 is an encoder-decoder based CPLM.
DeepseekCoder is a decoder-only model and we adopt \textit{deepseek-coder-1.3b-base} with 1.3B parameters.
CodeLlama is a decoder-only model based on Llama 2 and we adopt \textit{codellama-7b-base} with 7B parameters.
For CodeBERT and CodeT5, we pre-train them from scratch to generate target models, resulting CodeBERT with 125M parameters and CodeT5 with 220M parameters.
For larger CPLMs DeepseekCoder and CodeLlama, we continue training based on their released models to generate target models.

\subsubsection{\task Baselines}
We compare \ours with three \task baselines:
\begin{itemize}

    \item\textbf{FastText}~\cite{GraveMJB17}: It is a text classification library, which can be used to intuitively demonstrate whether there are distributional differences between member data and non-member data. We train FastText with the member data and the non-member data of $\mathcal{M}$, and use it to directly assess whether a code snippet is member or non-member data.

    \item\textbf{Perturbation}~\cite{ExtractLLM}: It perturbs a code snippet by converting case and calculates the L2 distance before and after the transformation to determine whether it is member data.

    \item\textbf{Perplexity}~\cite{ExtractLLM,abs-2101-05405,OhPKPK23}: It calculates the perplexity of an interested data record to determine whether it is member data. The intuition behind it is that the member data may have lower perplexity. Perplexity is commonly used for the CMI task.

\end{itemize}

\subsubsection{Data}
\label{Experiment:data}

For CodeBERT and CodeT5, we choose CSN\footnote{\url{https://github.com/github/CodeSearchNet}}~\cite{CodeSearchNet} dataset since their authors pre-train CodeBERT and CodeT5 over CSN.
CSN dataset contains over 6 million code snippets from open-source projects on GitHub, spanning six programming languages (Python, Java, JavaScript, Go, Ruby, and PHP).
Due to limited computational resource, we only use python code snippets of CSN.
Code snippets are associated with metadata such code description written in natural language.
We sample disjoint segments of CSN to pre-train target, shadow and calibration models. 
Specifically, we sample 100,000 data records for pre-training the target model, 50,000 for pre-training the shadow model and the calibration model, respectively. For testing, we sample 10,000 member data records and 10,000 non-member data records.

For DeepseekCoder and CodeLlama, we adopt Magicoder-Evol-Instruct-110k (MEI)\footnote{\url{https://huggingface.co/datasets/ise-uiuc/Magicoder-Evol-Instruct-110K}}~\cite{abs-2312-02120}. 
Noted that MEI is generated by GPT-4 and the data leakage issue can be avoided. 
Specifically, we sample 30,000 data records of MEI for training the target model, 20,000 for training the shadow model and the calibration model, respectively. For testing, we sample 5,000 member data records and 5,000 non-member data records.

\subsubsection{Environment and Hyper-Parameters}

\label{sec:app_env_hyper}

We run the experiments on a machine with two Intel(R) Xeon(R) Silver 4214R CPU @ 2.40GHz, 256 GB main memory and eight NVIDIA GeForce RTX 3090. 
We implement CodeBERT and CodeT5 following their original papers since only their pre-training implementations are not public available. 
For training CodeLlama, we utilize deepspeed\footnote{\url{https://github.com/microsoft/DeepSpeed}} and ZERO 1 optimization with cpu offload~\cite{RajbhandariRRH20}. 
We set the batch size to 64 and learning rate to 5e-5.
Other hyper-parameters are set according to original papers.

\subsection{Experimental Results}

\subsubsection{Overall Performance}

\begin{table}[t]
  \centering
  \caption{Performance of white-box and black-box inference. ``-'' indicates the values are almost zero.}
  \resizebox{0.98\columnwidth}{!}{
    \begin{tabular}{ccccc}
    \toprule
    \multirow{2}[2]{*}{Method} & \multirow{2}[2]{*}{AUC} & \multicolumn{3}{c}{TPR} \\
          &       & 0.01\%FPR & 0.10\%FPR & 1.00\%FPR \\
    \midrule
    FastText & 0.500 & - & - & - \\
    $\text{CodeBERT}_{perb}$ & 0.505 & 0.00\% & 0.12\% & 0.90\% \\
    $\text{CodeT5}_{perb}$ & 0.499 & 0.03\% & 0.15\% & 1.01\% \\
    $\text{DeepseekCoder}_{ppl}$ & 0.501 & 0.30\% & 0.30\% & 1.21\% \\
    $\text{CodeLlama}_{ppl}$ & 0.670 & 2.23\% & 2.23\% & 8.50\% \\
    \midrule
    $\text{CodeBERT}_{wb}$ & 0.603 & 0.06\% & 0.28\% & 2.36\% \\
    $\text{CodeT5}_{wb}$ & 0.869 & 0.04\% & 1.42\% & 12.16\% \\
    $\text{DeepseekCoder}_{wb}$ & 0.722 & 1.16\% & 5.05\% & 16.33\% \\
    $\text{CodeLlama}_{wb}$ & 0.980 & 22.57\% & 43.01\% & 83.80\% \\
    \midrule
    $\text{CodeBERT}_{bb}$ & 0.602 & 0.03\% & 0.26\% & 2.42\% \\
    $\text{CodeT5}_{bb}$ & 0.859 & 0.14\% & 1.65\% & 12.06\% \\
    $\text{DeepseekCoder}_{bb}$ & 0.721 & 1.08\% & 5.01\% & 16.33\% \\
    $\text{CodeLlama}_{bb}$ & 0.979 & 21.90\% & 41.07\% & 83.45\% \\
    \bottomrule
    \end{tabular}%
    }
  \label{tab:overall}%
\end{table}%

Tab.~\ref{tab:overall} provides the overall results. The abbreviations $bb$ and $wb$ stand for black-box \ours and white-box \ours, respectively. $perb$ and $ppl$ indicate Perturbation and Perplexity, respectively.

From Tab.~\ref{tab:overall}, we have the following findings:
\begin{itemize}
\item Member data and non-member data can not be easily separated according to code features (i.e., data distribution), as evidenced by the results of FastText: it achieves an AUC score close to 0.5.

\item \ours consistently shows superior performance than baselines FastText, Perturbation and Perplexity, showing the effectiveness of \ours. 

\item The AUC of white-box inference is not much higher than that of black-box inference. Hence, we can conclude that knowing the distribution of the training dataset of the target model has a relatively minor impact on the inference accuracy. \emph{In other words, \task in the black-box setting can be achieved with high accuracy}.

\item The AUC scores for CodeT5 (\textasciitilde0.8), DeepseekCoder (\textasciitilde0.7) and CodeLlama (\textasciitilde0.9) are much higher than that of CodeBERT (\textasciitilde0.6). A possible reason is that they have different model structures and parameter sizes. Recent works have found larger language models tend to over-memorized training data (member data) than smaller language models~\cite{TirumalaMZA22,CarliniIJLTZ23}, which may demonstrate why the AUC for CodeBERT, the smallest CPLM, is lowest.

\item  To investigate whether \ours suffers from high false positive rate, we show TPR under different FPR (0.01\%, 0.1\%, 1\%) in Tab.~\ref{tab:overall}. We can see that \ours overcomes the high FPR issue, a common problem in existing MI works~\cite{WatsonGCS22} since the TPR of \ours is much higher than TPR of the baselines under a low FPR. \ours shows much higher TPR on larger CPLMs DeepseekCoder and CodeLlama. The over-memorizing characteristic of larger language models~\cite{TirumalaMZA22,CarliniIJLTZ23} may be the reason.

\end{itemize}

\subsubsection{Impact of Data Characteristics}
\label{sec:data_char}

\begin{figure*}[t]
    \centering
    \includegraphics[width=1\linewidth]{images/ImpactOfCodeFeature.pdf}
    \caption{Impact of different code features.}
    \label{fig:ImpactOfCodeFeature}
\end{figure*}

Next, we study the impacts of different code characteristics on the inference results.
In other words, we are interested in the factors that affect how \ours makes membership status predictions.
We investigate three common code features:
\begin{itemize}

\item \textbf{Code Length:} Code length refers to the length of a code snippet. Longer code snippets can provide more information.

\item \textbf{Depth of AST:} The abstract syntax tree is an important feature that distinguishes code from natural language. The depth of the code abstract syntax tree may affect the inference results.

\item \textbf{Node Number of AST:} An AST node represents a fundamental component of the structure of a code snippet. Thus the number of AST node may affect the inference results.

\end{itemize}

Fig.~\ref{fig:ImpactOfCodeFeature} reports the distributions of the confidence scores of code snippets predicted by $\text{CodeBERT}_{wb}$ and $\text{CodeT5}_{wb}$. 
Due to page limit, we only show the results of CodeBERT and CodeT5.
First, we obtain the values for the three features of each code snippet. 
Next, we group the code data into intervals with equal length, and arrange intervals in ascending order based on the corresponding feature.
The x axis of Fig.~\ref{fig:ImpactOfCodeFeature} represents the intervals. 
The left y axis is the number of code samples in each interval and 
the right y axis denotes the confidence score. 
We normalize the scores to a range between 0 and 1 via min-max normalization.
The bar charts in Fig.~\ref{fig:ImpactOfCodeFeature} represents the 
number of samples in each group, while the line graphs show the average confidence scores of each group. Subsequently, we examine the scores assigned by the inference model to different groups and 
analyze whether any specific patterns emerge across the intervals.

From Fig.~\ref{fig:ImpactOfCodeFeature}, we can observe the long-tail distributions of code snippets grouped by the three code features. 
If we consider both confidence scores and number of code snippets for each interval, then we can find that, for CodeT5, the predicted confidence scores are positively correlated with code length, depth of AST, and number of AST nodes.

Differently, for CodeBERT, they are negatively correlated.
The differences are possibly caused by their model structures. For CodeT5, it is encoder-decoder structure, which consists of bidirectional attention and unidirectional attention mechanisms. When the sequence is long, tokens closer to the beginning of the sequence in the decoder receive more attention, leading to stronger training signals.

\begin{figure*}[ht]
  \centering
  \includegraphics[width=0.97\linewidth]{images/calibraion_compare.pdf}
  \caption{Impact of calibration.}
  \label{fig:calibration_compare}
\end{figure*}

\subsubsection{Analysis of Calibration Model} 

\begin{table}[t]
  \centering
  \caption{Impact of the calibration model (AUC).}
  \resizebox{0.87\linewidth}{!}{
    \begin{tabular}{ccccc}
    \toprule
    \multirow{2}[2]{*}{Method} & \multicolumn{2}{c}{White Box} & \multicolumn{2}{c}{Black Box} \\
          & w/ cal & w/o cal & w/ cal & w/o cal \\
    \midrule
    CodeBERT & 0.603 & 0.523 & 0.602 & 0.524 \\
    CodeT5 & 0.869 & 0.732 & 0.859 & 0.719 \\
    DeepseekCoder & 0.722 & 0.514 & 0.721 & 0.514 \\
    CodeLlama & 0.980 & 0.817 & 0.979 & 0.809 \\
    \bottomrule
    \end{tabular}%
  }
  \label{tab:caliberate}%
\end{table}%

\begin{table}[t]
  \centering
  \caption{Impact of Training Data Size (AUC).}
  \resizebox{0.75\linewidth}{!}{
    \begin{tabular}{cccc}
    \toprule

    \multirow{2}[2]{*}{Model} & \multicolumn{3}{c}{Data Number} \\
        & 20K & 10K & 5K \\

    \midrule
    $\text{DeepseekCoder}_{bb}$ & 0.721 & 0.696 & 0.657 \\
    \bottomrule
    \end{tabular}%
}
  \label{tab:dataset_size}%
\end{table}%

\begin{table}[t]
  \centering
  \caption{HumanEval (pass@1) of the target model.}
  \resizebox{0.65\linewidth}{!}{
    \begin{tabular}{ccccc}
    \toprule
    Method & Before & After \\
    \midrule
    DeepseekCoder & 0.34 & 0.36 \\
    CodeLlama & 0.30 & 0.42 \\
    \bottomrule
    \end{tabular}%
  }
  \label{tab:humaneval}%
\end{table}%

Tab.~\ref{tab:caliberate} displays the inference results with and without the calibration model. It is evident that the calibration model can significantly improve the \task performance. Specifically, for CodeBERT, it increases the AUC score by 0.08, and for CodeT5, by 0.14. For DeepseekCoder and CodeLlama, the effect of calibration on black-box inference is more significant. 
The calibration model is effective in both white-box inference and black-box inference. 

Fig.~\ref{fig:calibration_compare} further shows the improvements brought by the calibration model.
We bucketize the data w.r.t. code length in a similar way as Sec.~\ref{sec:data_char}.
Note that a higher AUC score indicates that the model can better separate member and non-member data, i.e., the gap between the confidence scores of member and non-member data is larger.
In Fig.~\ref{fig:calibration_compare} (a), the minimum score of member data with calibration is around 0.52 (the blue curve in Fig.~\ref{fig:calibration_compare} (a)), while the maximum score of non-member data with calibration is around 0.41 (the blue curve in Fig.~\ref{fig:calibration_compare} (b)). The gap with calibration is 0.11.
The minimum score of member data without calibration is around 0.62 (the orange curve in Fig.~\ref{fig:calibration_compare} (a)), while the maximum score of non-member data without calibration is around 0.59 (the orange curve in Fig.~\ref{fig:calibration_compare} (b)).
The gap without calibration is 0.03, which is smaller than the gap with calibration (0.11).
Hence, we can conclude that calibration increase the gap between member and non-member data, resulting in higher AUC.

\subsubsection{Impact of Training Data Size} 

To investigate how the size of training data for constructing \ours affects the performance, we report the effects of different data size on \ours over DeepseekCoder in black-box setting in Tab.~\ref{tab:dataset_size}.
Note that the results of $\text{DeepseekCoder}_{bb}$ in Tab.~\ref{tab:overall} is reported using default training data size 20K. 
From Tab.~\ref{tab:dataset_size}, we can see that, as the size of training data decreases, AUC of \ours declines.
However, when the training data size is small (5K), \ours can still achieve an AUC score of 0.657, showing that it can be effective even when not much training data is available.

\subsubsection{Effectiveness of Training Larger CPLMs} 

As we continue training with additional data over the pre-trained DeepseekCoder and CodeLlama to generate the target models, it is essential to investigate how the training affects the performance of the two large CPLMs.
Tab.~\ref{tab:humaneval} shows the performance of DeepseekCoder and CodeLlama on the HumanEval benchmark~\cite{abs-2107-03374}, a popular benchmark to evaluate code generation. 
We can that see that the pass@1 rate has increased after our training, indicating that the process of constructing target models positively affects DeepseekCoder and CodeLlama.

%% file: tex/conclusion.tex

\section{Conclusion}
\label{sec:conclusion}

In this paper, we study \task for authenticating data compliance in CPLMs and propose a framework \textsc{Buzzer} for inferring code membership. 
\textsc{Buzzer} achieves promising results on various CPLMs as shown in the experiments. 
\textsc{Buzzer} can serve as a \task tool and help protect the intellectual property rights.
In the future, we plan to further improve the generalization ability of \textsc{Buzzer} to make it more practical.
We will also explore the idea of this work on other multimodal pre-trained language models beyond CPLMs.

%% file: tex/limitations.tex

\section{Limitations}
\label{sec:limit}

We study \task using public code data that is not originally designed for this task.
In practice, the code data that their owners care about may not be publicly available, making it difficult to collect them for the study of \task.
For such cases, it is difficult to assess the performance of \ours based on the results reported in this paper.